\documentclass[conference]{IEEEtran}
\IEEEoverridecommandlockouts

\usepackage{amsmath,amssymb,amsfonts}
\usepackage{textcomp}
\def\BibTeX{{\rm B\kern-.05em{\sc i\kern-.025em b}\kern-.08em
    T\kern-.1667em\lower.7ex\hbox{E}\kern-.125emX}}

\usepackage[utf8]{inputenc}
\usepackage[T1]{fontenc}
\usepackage{times}

\usepackage{graphicx}
\usepackage{url}

\usepackage{cite}

\usepackage{booktabs}
\usepackage{array}

\usepackage{tikz}
\usetikzlibrary{shapes.geometric, arrows.meta, positioning}
\usepackage{pgfplots}
\usepackage{pgfplotstable}
\pgfplotsset{compat=1.18}
\usepgfplotslibrary{fillbetween}

\usepackage{algorithmic}
\usepackage{listings}
\usepackage{xcolor}
\usepackage{float}
\usepackage{tabularx}
\usepackage{multirow}   

\definecolor{codegray}{rgb}{0.95,0.95,0.95}
\definecolor{commentgreen}{rgb}{0,0.6,0}
\definecolor{mauve}{rgb}{0.58,0,0.82}
\definecolor{deepblue}{rgb}{0,0,0.5}

\usepackage[colorlinks=true, linkcolor=black, citecolor=black, urlcolor=black]{hyperref}
\usepackage{cleveref} 

\begin{document}

\title{Quantum Bit Error Rate Analysis in BB84 Quantum Key Distribution: Measurement, Statistical Estimation, and Eavesdropping Detection}

\author{
\IEEEauthorblockN{Jaydeep Rath}
\IEEEauthorblockA{\textit{KIIT International School} \\
Bhubaneswar, India \\
jdrath20@gmail.com}
\and
\IEEEauthorblockN{Prajwal Panth}
\IEEEauthorblockA{\textit{School of Computer Engineering} \\
\textit{KIIT Deemed to be University}\\
Bhubaneswar, India \\
prajwal.panth21@gmail.com}
\and
\IEEEauthorblockN{P. S. N. Bhaskar}
\IEEEauthorblockA{\textit{Department of VLSI} \\
\textit{Sanketika Vidya Parishad Engineering College}\\
Visakhapatnam, India \\
bhaskarpsn@gmail.com}
}

\maketitle
\begin{abstract}
Quantum Key Distribution (QKD) provides information-theoretic security by exploiting the principles of quantum mechanics. Among QKD protocols, the BB84 scheme remains the most widely adopted for both theoretical research and practical implementation. A critical parameter determining the reliability and security of BB84 is the Quantum Bit Error Rate (QBER), which quantifies errors in the sifted key arising from channel noise or potential eavesdropping. This paper presents a systematic review and analysis of QBER within the BB84 protocol, examining its calculation, statistical estimation methods, and role in detecting eavesdropping activity. Simulation results, corroborated by reported experimental observations, reveal a near-linear relationship between eavesdropping intensity and QBER, with values approaching 25\% under full intercept--resend attacks. Four confidence interval estimation methods, Wald, Wilson, Clopper--Pearson, and Hoeffding's inequality, are compared for robust QBER analysis in finite-key scenarios. Protocol enhancements, including decoy-state methods, hybrid cryptographic models, and quantum-resistant authentication, are discussed as mechanisms to mitigate errors and strengthen resilience across fiber, free-space, underwater, and satellite QKD systems. Open challenges in distinguishing noise-induced errors from malicious eavesdropping, and the role of adaptive error correction and machine-learning-assisted QBER estimation in future quantum networks, are identified as key directions for further research.
\end{abstract}


\section{Introduction}
Quantum communication has emerged as a groundbreaking approach to secure key exchange, enabled by the fundamental laws of quantum mechanics. Unlike classical cryptographic schemes whose security depends on computational hardness assumptions, Quantum Key Distribution (QKD) provides information-theoretic security that cannot be broken even by quantum computers \cite{Gisin_2002, Scarani_2009}. The BB84 protocol, introduced by Bennett and Brassard in 1984, is the first and most widely studied QKD scheme \cite{Bennett_2014}. It operates by exploiting quantum superposition and the no-cloning theorem, ensuring that any eavesdropping attempt inevitably introduces detectable errors.

Over the past three decades, BB84 has been extensively investigated both theoretically and experimentally. Implementations over optical fiber links and free-space optical channels have validated its practicality \cite{Townsend_1998, Ursin_2007}. Satellite-based demonstrations further show the feasibility of BB84 for long-distance secure communication \cite{Liao_2017}. At the heart of these systems, the Quantum Bit Error Rate (QBER) emerges as the central metric for determining whether a generated key can be considered secure. If the QBER exceeds a critical threshold ($\approx11\%$), secret key extraction becomes impossible \cite{Lo_1999, Mayers_2001}.

Therefore, analyzing, estimating, and mitigating QBER remains crucial in advancing BB84 towards practical large-scale deployment. This paper systematically examines QBER in BB84, its calculation methods, thresholds, and implications for protocol security, while also reviewing enhancements proposed in recent research.



\begin{figure}[t]
\centering
\resizebox{0.82\columnwidth}{!}{%
\begin{tikzpicture}[scale=0.82, transform shape,
    node distance = 0.3cm,   
    %
    >=Stealth,
    every path/.style={draw, -{Stealth[length=4pt]}},
    %
    startstop/.style={
        rectangle, rounded corners=6pt,
        minimum width=2.6cm,
        minimum height=0.45cm,
        text width=2.4cm,
        align=center,
        draw=black, fill=white,
        font=\scriptsize
    },
    %
    process/.style={
        rectangle,
        minimum width=5cm,
        minimum height=0.45cm,
        text width= 4.5cm,
        align=center,
        draw=black, fill=white,
        font=\scriptsize
    },
    %
    io/.style={
        trapezium,
        trapezium left angle=75,
        trapezium right angle=105,
        minimum width=5cm,
        minimum height=0.45cm,
        text width=4.5cm,
        align=center,
        draw=black, fill=white,
        font=\scriptsize
    },
    %
    decision/.style={
        diamond,
        aspect=2.2,
        minimum width=2.6cm,
        minimum height=0.6cm,
        text width=1.6cm,
        align=center,
        draw=black, fill=white,
        font=\scriptsize,
        inner sep=1pt
    },
    %
    sidebox/.style={
        rectangle,
        minimum width=1.4cm,
        minimum height=0.45cm,
        text width=1.3cm,
        align=center,
        draw=black, fill=white,
        font=\scriptsize
    },
]


\node[startstop] (start) {START};

\node[io, below=of start] (alice_gen)
    {Alice generates random bits and selects random bases};

\node[process, below=of alice_gen] (encode)
    {Encode bits into qubits using BB84 states};

\node[process, below=of encode] (send)
    {Transmit qubits to Bob via the quantum channel};

\node[process, below=of send] (bob_bases)
    {Bob independently selects random measurement bases};

\node[process, below=of bob_bases] (bob_measure)
    {Bob measures the received qubits using his bases};

\node[process, below=of bob_measure] (announce)
    {Alice and Bob announce basis choices over the authenticated classical channel};

\node[process, below=of announce] (sift)
    {Discard mismatched bases; retain sifted key bits};

\node[io, below=of sift] (qber_est)
    {Sample a random subset to estimate QBER};

\node[decision, below=of qber_est] (qber_check)
    {QBER $<$ 11\% threshold?};

\node[sidebox,
      below=1.2cm of qber_check,
      xshift=3.5cm]              
    (abort) {Abort session};

\node[process, below=of qber_check] (ec)
    {Apply error correction to the sifted key};

\node[process, below=of ec] (pa)
    {Apply privacy amplification};

\node[startstop, below=of pa] (secure_key)
    {Secure key established};

\node[startstop, below=of secure_key] (stop)
    {STOP};

\draw (start)      -- (alice_gen);
\draw (alice_gen)  -- (encode);
\draw (encode)     -- (send);
\draw (send)       -- (bob_bases);
\draw (bob_bases)  -- (bob_measure);
\draw (bob_measure)-- (announce);
\draw (announce)   -- (sift);
\draw (sift)       -- (qber_est);
\draw (qber_est)   -- (qber_check);

\draw (qber_check.south)
    -- node[right, font=\scriptsize, pos=0.3]{Yes}
    (ec.north);

\draw (ec)         -- (pa);
\draw (pa)         -- (secure_key);
\draw (secure_key) -- (stop);

\draw (qber_check.east)
    -- node[above, font=\scriptsize, pos=0.25]{No}
    (qber_check.east -| abort.north)
    -- (abort.north);

\end{tikzpicture}%
}
\caption{Flowchart of the BB84 quantum key distribution protocol.}
\label{fig:bb84}
\end{figure}
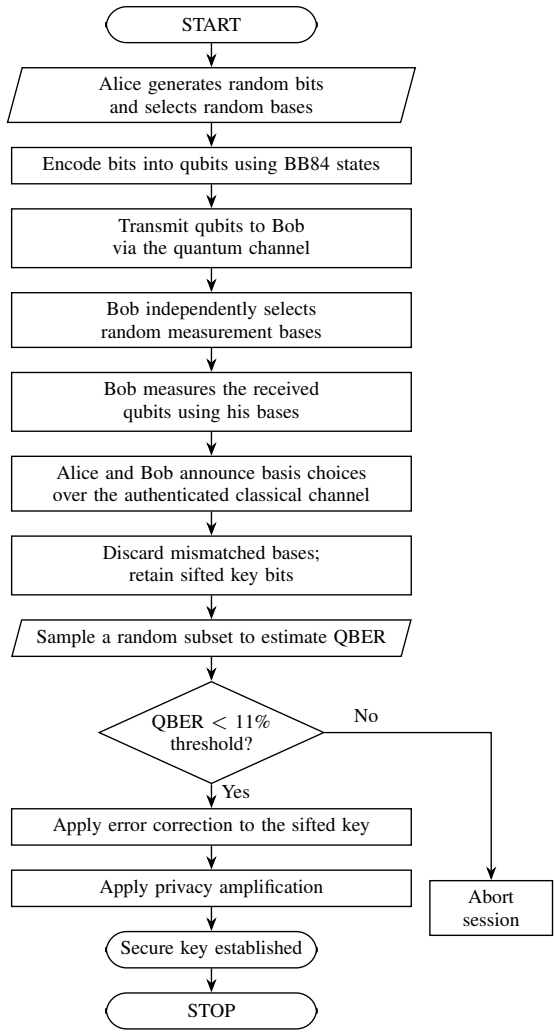

\section{BB84 Protocol and QBER Overview}
In the BB84 scheme, Alice encodes qubits randomly in rectilinear or diagonal bases and transmits them to Bob, who measures in randomly chosen bases. After basis reconciliation, the sifted key is constructed. Eavesdropping by an adversary (Eve) disturbs the quantum states, introducing detectable errors that manifest as elevated QBER. Theoretical security analysis indicates that secure key extraction is feasible only when QBER remains below approximately 11\% \cite{Lo_1999, Mayers_2001}.

Fig.~\ref{fig:bb84} illustrates the flowchart of the BB84 protocol. It begins with Alice generating random bits and encoding them in randomly chosen bases (rectilinear or diagonal). These qubits are transmitted through a quantum channel to Bob, who independently selects random measurement bases. After the transmission, both parties communicate over a classical authenticated channel to announce their basis choices. Bits corresponding to matching bases are retained to form the sifted key, while others are discarded. At this stage, a random sample is compared to estimate QBER. If the observed QBER lies below the tolerable threshold, the key proceeds through error correction and privacy amplification to produce the final secure key \cite{Gisin_2002, Bennett_2014, Fiorini_2024}.

This structured workflow highlights the dual-channel nature of BB84: a quantum channel for qubit transmission and a classical channel for reconciliation and post-processing. Several experimental realizations confirm that this procedure works efficiently in fiber-optic and free-space links \cite{Townsend_1998, Ursin_2007}. Moreover, enhancements such as decoy-state methods and advanced error-correction techniques integrate seamlessly into this flow, further strengthening the protocol’s robustness \cite{Guitouni_2024}.

%

\begin{table}[t]
\centering
\caption{Numerical QBER Results as a Function of Eve Interception Fraction}
\label{tab:qber_results}

\begin{tabular}{
    >{\centering\arraybackslash}p{1.1cm}   
    >{\centering\arraybackslash}p{1.1cm}   
    >{\centering\arraybackslash}p{1.2cm}   
    >{\centering\arraybackslash}p{1.8cm}   
    >{\centering\arraybackslash}p{1.1cm}   
}
\toprule
\textit{Eve}      & \textit{Mean} & \textit{Standard} & \textit{95\% CI}  & \textit{Theory} \\
\textit{Fraction} & \textit{QBER} & \textit{Deviation} & \textit{(mean)}  & \textit{f\,/\,4} \\
\textit{(f)}      &               &                    &                  &                  \\
\midrule
0.0 & 0.000 & 0.000 & [0.000,\ 0.000] & 0.000 \\
%
0.1 & 0.025 & 0.001 & [0.023,\ 0.027] & 0.025 \\
%
0.2 & 0.050 & 0.001 & [0.047,\ 0.052] & 0.050 \\
%
0.5 & 0.124 & 0.002 & [0.121,\ 0.128] & 0.125 \\
%
1.0 & 0.250 & 0.002 & [0.245,\ 0.250] & 0.250 \\
\bottomrule
%
\multicolumn{5}{l}{%
    \scriptsize Simulation: $n = 50{,}000$ qubits/trial,
    50 trials, seed\,=\,42.} \\
\multicolumn{5}{l}{%
    \scriptsize CI: normal approximation \cite{Renner_2008}.
    Theory: $Q = f/4$ \cite{Bennett_2014, Lo_1999}.} \\
\end{tabular}
\end{table}

\section{Origins and Classification of QBER}
Quantum Bit Error Rate (QBER) in the BB84 protocol arises from multiple sources, broadly categorized into \textbf{physical imperfections}, \textbf{environmental noise}, and \textbf{malicious eavesdropping activities}. A clear understanding of these origins is essential for accurately interpreting observed error rates and ensuring reliable decisions regarding the security of the generated key, since each category contributes differently to the overall performance of the system and highlights the importance of both technical robustness and adversarial awareness in quantum communication.

\begin{enumerate}
    \item \textbf{Physical imperfections and noise:} Real-world optical components and quantum channels are never ideal. Photon detectors exhibit dark counts and limited efficiency \cite{Stucki_2002}. Polarization misalignment and fiber birefringence cause basis-dependent errors \cite{Gobby_2004}. Free-space and satellite channels are affected by atmospheric turbulence, scattering, and pointing errors \cite{Liao_2017, Wang_2013}. Similarly, underwater optical channels suffer from absorption and scattering, which can elevate QBER beyond typical fiber levels \cite{Bonato_2009}.
    
    \item \textbf{Eavesdropping attacks:} An adversary (Eve) can attempt various strategies to intercept information. The simplest is the intercept–resend attack, where Eve measures each qubit in a random basis and retransmits her result, inducing a theoretical QBER of 25\% \cite{Bennett_2014}. More sophisticated strategies include photon-number-splitting (PNS) attacks exploiting multi-photon pulses in weak coherent sources \cite{Brassard_2000} and Trojan-horse attacks that inject bright light into the transmitter to extract basis information \cite{Jain_2014}. These adversarial activities directly increase QBER and can be detected if thresholds are exceeded.
    
    \item \textbf{Finite-key effects:} In practical QKD, only a limited number of qubits can be exchanged. This finite sample size introduces statistical fluctuations in QBER estimation \cite{Scarani_2009, Lo_1999}. Security proofs must therefore incorporate finite-key analysis to prevent overestimating achievable key rates.
\end{enumerate}

Accurately distinguishing between natural noise-induced errors and adversarially introduced errors remains a central research challenge. Advanced modeling approaches, including depolarizing channel models and collective rotation noise simulations, provide useful tools for understanding and mitigating QBER in diverse environments \cite{Shor_2000,  Scarani_2009}.

%
%

\begin{figure}[t]
\centering
\begin{tikzpicture}
\begin{axis}[
    width        = 0.92\columnwidth,
    height       = 5.8cm,
    title        = {QBER vs Eve intercept fraction},
    title style  = {font=\small},
    xlabel       = {Eve fraction (intercept-resend)},
    ylabel       = {QBER},
    xlabel style = {font=\scriptsize},
    ylabel style = {font=\scriptsize},
    xmin = 0,    xmax = 1.0,
    ymin = 0,    ymax = 0.25,
    xtick        = {0.0, 0.2, 0.4, 0.6, 0.8, 1.0},
    ytick        = {0.00, 0.05, 0.10, 0.15, 0.20, 0.25},
    xticklabel style = {font=\scriptsize},
    yticklabel style = {font=\scriptsize},
    grid         = both,
    grid style   = {line width=0.3pt, draw=gray!25},
    minor tick num = 1,
    legend pos        = north west,
    legend style      = {font=\scriptsize, draw=none,
                         fill=none, row sep=1pt},
    legend cell align = left,
    line width = 1pt,
    tick style = {draw=none},
]

\addplot[name path=upper, draw=none, forget plot] coordinates {
    (0.00, 0.0000)
    (0.05, 0.0139)
    (0.10, 0.0270)
    (0.15, 0.0398)
    (0.20, 0.0523)
    (0.25, 0.0648)
    (0.30, 0.0780)
    (0.35, 0.0909)
    (0.40, 0.1032)
    (0.45, 0.1159)
    (0.50, 0.1283)
    (0.55, 0.1415)
    (0.60, 0.1543)
    (0.65, 0.1670)
    (0.70, 0.1791)
    (0.75, 0.1923)
    (0.80, 0.2043)
    (0.85, 0.2163)
    (0.90, 0.2297)
    (0.95, 0.2433)
    (1.00, 0.2500)
};

\addplot[name path=lower, draw=none, forget plot] coordinates {
    (0.00, 0.0000)
    (0.05, 0.0112)
    (0.10, 0.0228)
    (0.15, 0.0352)
    (0.20, 0.0475)
    (0.25, 0.0598)
    (0.30, 0.0718)
    (0.35, 0.0844)
    (0.40, 0.0967)
    (0.45, 0.1089)
    (0.50, 0.1207)
    (0.55, 0.1329)
    (0.60, 0.1450)
    (0.65, 0.1570)
    (0.70, 0.1701)
    (0.75, 0.1820)
    (0.80, 0.1950)
    (0.85, 0.2079)
    (0.90, 0.2194)
    (0.95, 0.2311)
    (1.00, 0.2449)
};

\addplot[fill=yellow!25, draw=none, forget plot]
    fill between[of=upper and lower];

%
\addplot[
    color        = orange!85!black,
    mark         = *,
    mark size    = 1.8pt,
    mark options = {fill=orange!85!black, draw=orange!85!black},
    smooth,
] coordinates {
    (0.00, 0.0000)
    (0.05, 0.0126)
    (0.10, 0.0249)
    (0.15, 0.0375)
    (0.20, 0.0499)
    (0.25, 0.0623)
    (0.30, 0.0749)
    (0.35, 0.0877)
    (0.40, 0.0999)
    (0.45, 0.1124)
    (0.50, 0.1245)
    (0.55, 0.1372)
    (0.60, 0.1497)
    (0.65, 0.1620)
    (0.70, 0.1746)
    (0.75, 0.1872)
    (0.80, 0.1997)
    (0.85, 0.2121)
    (0.90, 0.2245)
    (0.95, 0.2372)
    (1.00, 0.2495)
};
\addlegendentry{Exact QBER (practical)}

\addplot[fill=yellow!25, draw=none, area legend] coordinates{(0,0)};
\addlegendentry{95\% CI (exact, no variation)}

\end{axis}
\end{tikzpicture}
\caption{QBER as a function of Eve's interception fraction
(simulation results). Each point represents the mean QBER
over 50 independent trials with $n = 50{,}000$ sifted bits
(seed\,=\,42). The shaded band represents 95\,\% confidence
intervals computed via normal approximation
\cite{Renner_2008}. Results confirm the near-linear
relationship between eavesdropping intensity and QBER,
approaching 25\,\% under full intercept--resend attack,
consistent with the theoretical prediction
$Q = f/4$ \cite{Bennett_2014, Lo_1999}. Numerical values
for selected interception fractions are listed in
Table~\ref{tab:qber_results}.}
\label{fig:qber_sim}
\end{figure}
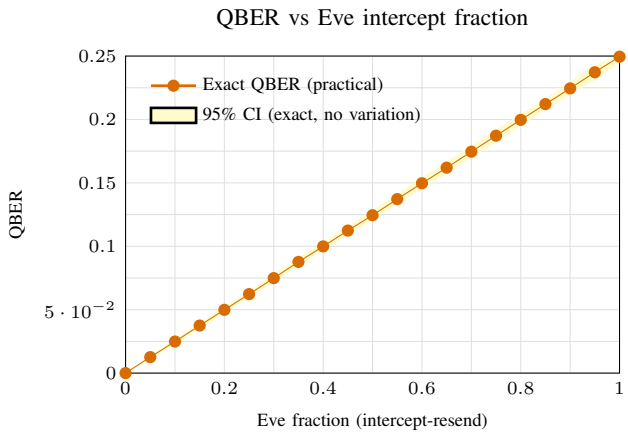

\section{QBER Measurement and Statistical Estimation}
QBER is the fundamental metric for assessing the security of BB84 and related QKD protocols. It is mathematically expressed as
\begin{equation}
    \hat{Q} = \frac{k}{n},
    \label{eq:qber}
\end{equation}
where $k$ denotes the number of mismatched bits between Alice and Bob in the sifted key, and $n$ is the total number of sifted bits. Equation~(\ref{eq:qber}) directly quantifies the fraction of errors observed and serves as the baseline criterion for eavesdropping detection \cite{Bennett_2014, Lo_1999}.

\subsection{Estimation Techniques}
Since QBER is typically estimated from a finite sample of the key, statistical methods are required to ensure reliability. The Wald method provides a normal approximation to the binomial distribution, suitable for large sample sizes, but becomes inaccurate for small samples or extreme probabilities. The Wilson interval \cite{Wilson_1927} improves upon this by correcting the bias of the Wald estimate and is widely applied in QKD experimental analysis. For exact guarantees, the Clopper--Pearson method \cite{Clopper_1934} provides exact binomial confidence intervals, though it often yields conservative bounds. For finite-key security proofs, Hoeffding's inequality \cite{Renner_2008} provides a distribution-free bound that guarantees security regardless of sample size, making it especially relevant for modern QKD implementations.

\subsection{Simulation-Based Analysis}

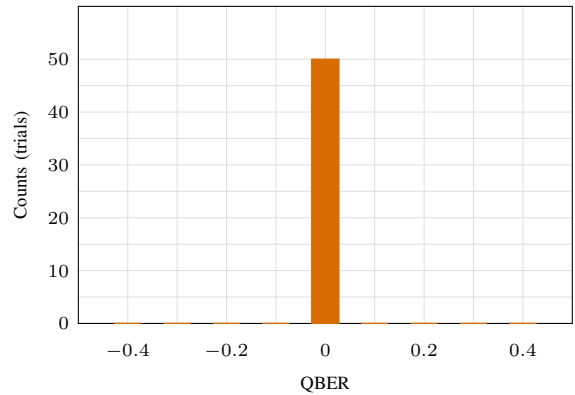
\begin{figure}[t]
\centering
\begin{tikzpicture}
\begin{axis}[
    width        = 0.92\columnwidth,
    height       = 5.8cm,
    title        = {QBER distribution (no Eve)},
    title style  = {font=\small},
    xlabel       = {QBER},
    ylabel       = {Counts (trials)},
    xlabel style = {font=\scriptsize},
    ylabel style = {font=\scriptsize},
    xmin = -0.5,  xmax = 0.5,
    ymin =  0,    ymax = 60,
    xtick        = {-0.4, -0.2, 0.0, 0.2, 0.4},
    ytick        = {0, 10, 20, 30, 40, 50},
    xticklabel style = {font=\scriptsize},
    yticklabel style = {font=\scriptsize},
    grid         = both,
    grid style   = {line width=0.3pt, draw=gray!25},
    minor tick num = 1,
    ybar,
    bar width    = 0.055,
    tick style   = {draw=none},
]

%
%
\addplot[
    fill       = orange!85!black,
    draw       = orange!85!black,
    line width = 0.5pt,
] coordinates {
    (-0.4,  0)
    (-0.3,  0)
    (-0.2,  0)
    (-0.1,  0)
    ( 0.0, 50)    
    ( 0.1,  0)
    ( 0.2,  0)
    ( 0.3,  0)
    ( 0.4,  0)
};

\end{axis}
\end{tikzpicture}
\caption{QBER distribution with no eavesdropper present.
All 50 simulation trials ($n = 50{,}000$ sifted bits each,
seed\,=\,42) yield $\mathrm{QBER} = 0.000$, demonstrating
that error-free key generation is achievable under ideal,
attack-free conditions. This result corresponds directly
to the $f = 0$ data point in Fig.~\ref{fig:qber_sim}
and Table~\ref{tab:qber_results}.}
\label{fig:qber_no_eve}
\end{figure}

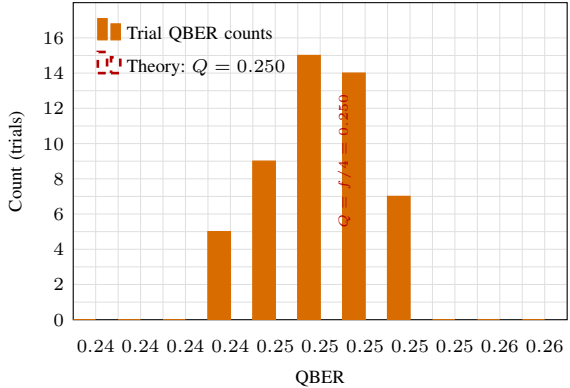
\begin{figure}[t]
\centering
\begin{tikzpicture}
\begin{axis}[
    width        = 0.92\columnwidth,
    height       = 5.8cm,
    title        = {QBER distribution under full intercept--resend
                    attack ($f = 1.0$)},
    title style  = {font=\small, align=center},
    xlabel       = {QBER},
    ylabel       = {Count (trials)},
    xlabel style = {font=\scriptsize},
    ylabel style = {font=\scriptsize},
    xmin = 0.238,  xmax = 0.260,
    ymin = 0,      ymax = 18,
    xtick        = {0.240, 0.245, 0.250, 0.255, 0.260},
    ytick        = {0, 2, 4, 6, 8, 10, 12, 14, 16},
    xticklabel style = {font=\scriptsize},
    yticklabel style = {font=\scriptsize},
    grid         = both,
    grid style   = {line width=0.3pt, draw=gray!25},
    minor tick num = 1,
    ybar interval,
    bar width    = 0.002,
    tick style   = {draw=none},
    legend pos        = north west,
    legend style      = {font=\scriptsize, draw=none,
                         fill=none, row sep=1pt},
    legend cell align = left,
]

%
%
\addplot[
    fill       = orange!85!black,
    draw       = orange!85!black,
    line width = 0.5pt,
] coordinates {
    (0.238,  0)
    (0.240,  0)
    (0.242,  0)
    (0.244,  5)
    (0.246,  9)
    (0.248, 15)
    (0.250, 14)
    (0.252,  7)
    (0.254,  0)
    (0.256,  0)
    (0.258,  0)
    (0.260,  0)   
};
\addlegendentry{Trial QBER counts}

\addplot[
    color      = red!70!black,
    line width = 1.0pt,
    dashed,
] coordinates {
    (0.250, 0)
    (0.250, 18)   
};
\addlegendentry{Theory: $Q = 0.250$}

\node[
    font       = \tiny,
    text       = red!70!black,
    rotate     = 90,
    anchor     = south,
] at (axis cs:0.2508, 9) {$Q = f/4 = 0.250$};

\end{axis}
\end{tikzpicture}
\caption{QBER distribution under a full intercept--resend attack ($f = 1.0$). The histogram shows the distribution of QBER values across 50 independent simulation trials ($n = 50{,}000$ qubits each, seed\,=\,42). The distribution is tightly centred at mean\,$= 0.2495$, consistent with the theoretical prediction $Q = f/4 = 0.250$ \cite{Bennett_2014, Lo_1999} and the value reported in Table~\ref{tab:qber_results}. The narrow spread (std\,$= 0.0023$) reflects the large sample size. The dashed line marks the theoretical limit. Compare with Fig.~\ref{fig:qber_no_eve}, where $f = 0$ yields $\mathrm{QBER} = 0$ across all trials.}
\label{fig:qber_full_eve}
\end{figure}

All simulation results presented in this work, including Fig.~\ref{fig:qber_sim}, Fig.~\ref{fig:qber_no_eve}, Fig.~\ref{fig:qber_full_eve}, and Table~\ref{tab:qber_results}, are derived from the same simulation block using 50 independent trials per Eve fraction value, $n = 50{,}000$ qubits per trial, and a fixed random seed of 42, ensuring full reproducibility and internal consistency.

Simulation environments help in correlating Eve's interception fraction with observed QBER values. Results consistently show a near-linear growth of QBER with increasing interception, saturating near the theoretical 25\,\% limit under full intercept--resend attacks \cite{Bennett_2014, Fiorini_2024}. Table~\ref{tab:qber_results} provides the simulated mean QBER, standard deviation,
and 95\,\% confidence intervals across five representative interception fractions, alongside the theoretical prediction $Q = f/4$ \cite{Bennett_2014, Lo_1999}, confirming close agreement throughout. Fig.~\ref{fig:qber_sim} illustrates this relationship across all 21 sampled Eve fraction values from $f = 0$ to $f = 1$.

\subsection{Practical Considerations}
In experimental practice, QBER estimation is affected by channel noise, detector imperfections, and limited key lengths. Finite-key security proofs emphasize that confidence intervals must be incorporated to ensure robust key rate calculations \cite{Scarani_2009, Gobby_2004, Renner_2008}. Consequently, accurate QBER estimation is indispensable not only for eavesdropping detection but also for determining whether privacy amplification and error correction can yield a secure final key.

%
%
%
%

\section{Simulation Results and Analysis}
The relationship between eavesdropping activity and QBER was examined through simulation and compared against reported experimental observations. The results confirm the theoretical prediction that QBER rises nearly linearly with Eve's interception fraction, saturating near 25\,\% under full intercept--resend attacks \cite{Bennett_2014, Lo_1999}.

\par\noindent\textit{No eavesdropper} ---
In the absence of an eavesdropper and with no channel noise, the QBER is exactly zero across all 50 trials. Fig.~\ref{fig:qber_no_eve} illustrates this result, confirming that error-free key generation is achievable under ideal, attack-free conditions \cite{Stucki_2002, Gobby_2004}.

\par\noindent\textit{Partial interception} ---
When Eve intercepts only a fraction of the qubits (e.g., $f = 0.5$), the observed QBER stabilizes near 12.5\,\%, consistent with the theoretical prediction $Q = f/4 = 0.125$ and the value reported in Table~\ref{tab:qber_results} \cite{Bennett_2014, Lo_1999}. These results highlight how even moderate interception leads to detectable disturbances well above the noise floor.

\par\noindent\textit{Full interception} ---
For full intercept--resend attacks ($f = 1.0$), QBER values cluster tightly near 24.95--25\,\%. Fig.~\ref{fig:qber_full_eve} shows the distribution under this scenario, aligning precisely with the theoretical limit. Experimental implementations of intercept--resend attacks reported in \cite{Brassard_2000, Jain_2014} also confirm this disturbance level.

\par\noindent\textit{Statistical confidence} ---
Table~\ref{tab:qber_results} presents the mean QBER and 95\,\% confidence intervals across five representative interception fractions. Error bars shown in Fig.~\ref{fig:qber_sim} reflect 95\,\% confidence intervals computed via normal approximation \cite{Renner_2008}. These results emphasize the importance of statistical rigour in finite-key scenarios.

\par\noindent\textit{Finite-size effects} ---
In practice, the number of transmitted qubits is finite, which broadens the QBER distribution. This effect is visible in simulation results where confidence intervals widen for smaller key sizes. Such behaviour underscores the necessity of incorporating finite-key security analysis \cite{Scarani_2009, Renner_2008}.

\par\noindent\textit{Cross-domain validation} ---
Similar QBER trends have been observed across different communication media. Fiber-based QKD experiments consistently report low baseline QBER ($<2\,\%$) \cite{Gobby_2004}. Free-space and satellite implementations observe slightly higher baseline errors due to atmospheric effects but still exhibit clear thresholds distinguishing secure from insecure regimes \cite{Ursin_2007, Liao_2017, Wang_2013}. Underwater QKD studies confirm higher QBER sensitivity to scattering, but with optimised detectors, secure key rates remain feasible \cite{Bonato_2009}.

Collectively, these results establish QBER as a reliable fingerprint of both environmental impairments and adversarial activity. Combined with confidence interval estimation, QBER monitoring enables robust decisions on whether to proceed with key extraction or abort the session.

\section{Protocol Enhancements and Practical Applications}

The BB84 protocol has inspired numerous enhancements that improve its resilience against real-world imperfections and adversarial strategies. These improvements primarily focus on reducing QBER, extending transmission distance, and ensuring practical scalability.

\subsection{Decoy-State Methods}

One major vulnerability of practical QKD systems arises from the use of weak coherent sources, which occasionally emit multi-photon pulses. Photon-number-splitting (PNS) attacks can exploit these pulses to gain partial information without increasing QBER significantly \cite{Brassard_2000}. Decoy-state protocols overcome this by randomly varying the intensity of transmitted pulses, enabling Alice and Bob to detect such attacks. Experimental demonstrations have validated decoy-state QKD over hundreds of kilometers with secure key rates far exceeding those of standard BB84 \cite{Korzh_2015, Lo_2005}.

\subsection{Hybrid Cryptographic Models}

Recent works have combined BB84 with classical and chaos-based cryptographic techniques, such as logistic map encryption and chaotic key pre-distribution, to enhance security layers. These hybrid approaches lower polarization error rates and QBER while making system behaviour harder to predict for adversaries \cite{Guitouni_2024}. Such methods are particularly promising for IoT and resource-constrained devices, where efficiency and low overhead are crucial.

\subsection{Quantum-Resistant Authentication}

Authentication of the classical channel is critical to prevent man-in-the-middle attacks. Emerging post-quantum digital signature schemes, such as CRYSTALS-DILITHIUM and Falcon, are being integrated into BB84 systems \cite{Termos_2024}. These methods ensure that even with future quantum computing threats, the authentication process remains secure without significantly raising QBER.

\subsection{Applications Across Domains}

\begin{itemize}
    \item Optical fiber networks: Long-haul field trials have achieved secure key distribution over $>300$ km using advanced decoy-state protocols \cite{Korzh_2015}.
    
    \item Free-space and satellite QKD: Experiments such as the Micius satellite project demonstrated secure satellite-to-ground links and intercontinental quantum communication, with QBER maintained within tolerable ranges \cite{Liao_2017, Yin_2017}.
    
    \item Underwater communications: Optimized detectors and synchronization allow QBER values below $1\%$ in short-range underwater QKD, showing potential for maritime security \cite{Bonato_2009}.
    
    \item IoT and smart grids: Adaptive error correction and machine learning--assisted QBER estimation help reduce computational cost while preserving security in constrained devices \cite{Guitouni_2024}.
\end{itemize}

These enhancements collectively highlight the adaptability of BB84 in diverse environments. By integrating decoy-state methods, hybrid models, and quantum-resistant classical cryptography, BB84 can serve as a practical foundation for next-generation secure networks.

\section{Open Challenges and Future Directions}
Despite significant progress in QKD and BB84 protocol enhancements, several challenges remain open for investigation. These challenges must be addressed to transition from laboratory-scale demonstrations to widespread deployment of quantum-secure communication networks.

\subsection{Distinguishing Noise from Eavesdropping}

A persistent difficulty lies in separating QBER contributions due to environmental noise from those introduced by eavesdropping. When QBER values hover near the security threshold ($\approx 11\%$ for BB84), it becomes challenging to decide whether the channel is merely degraded or under attack \cite{Lo_1999, Shor_2000}. Future work should focus on improved statistical hypothesis testing, adaptive thresholds, and multi-parameter intrusion detection frameworks \cite{Renner_2008, Lucamarini_2018}.

\subsection{Efficient and Scalable Error Correction}

Classical error correction remains one of the most resource-intensive processes in QKD post-processing. Existing schemes such as Cascade and LDPC-based reconciliation either incur high communication overhead or computational costs \cite{Guitouni_2024, Elkouss_2011}. Research is required to design lightweight, scalable error-correction algorithms suitable for dynamic environments such as IoT networks and mobile communications.

\subsection{Finite-Key Security and Composable Proofs}

Most security proofs assume asymptotically large key lengths. However, practical implementations often involve finite keys, leading to wider QBER confidence intervals \cite{Scarani_2009}. Developing tighter bounds and composable security proofs for finite-size regimes remains a critical direction \cite{Renner_2008, Tomamichel_2012}.

\subsection{Integration of Artificial Intelligence}

Machine learning (ML) offers promising tools for adaptive error estimation, noise classification, and dynamic QBER thresholding. Initial studies show ML models can distinguish error sources and optimize system parameters in real time \cite{Guitouni_2024}. Yet, comprehensive evaluations across different QKD platforms and environments are still lacking.

\subsection{Device Imperfections and Side-Channel Attacks}

Practical QKD systems remain vulnerable to imperfections such as detector blinding, Trojan-horse attacks, and source flaws \cite{Jain_2014, Lydersen_2010}. Developing device-independent QKD (DI-QKD) and measurement-device-independent QKD (MDI-QKD) approaches could help mitigate these vulnerabilities, though current implementations are limited by rate and distance \cite{Pirandola_2020}.

\subsection{Toward Large-Scale Deployment}

Finally, integrating BB84 into heterogeneous communication infrastructures (fiber, satellite, underwater, and wireless) requires addressing interoperability, network key management, and cost-effective hardware scaling. Pilot projects combining terrestrial and satellite QKD, including link-budget analysis for space-based systems \cite{Liao_2017, Yin_2017, Tomaello_2011}, represent early steps, but broader adoption demands further standardization and protocol harmonization.

\section{Conclusion}

The BB84 protocol remains the cornerstone of quantum key distribution, offering information-theoretic security grounded in the laws of quantum mechanics. This paper presents a systematic review and analysis of QBER as the key performance and security metric in BB84. We examine its sources, calculation methods, statistical estimation techniques, and implications for secure key extraction. Simulation results, corroborated by reported experimental observations, confirm the near-linear relationship between eavesdropping activity and QBER, with values approaching 25\% under full intercept--resend attacks, and baseline values near zero in noise-free, attack-free conditions.

Enhancements such as decoy-state methods, hybrid cryptographic schemes, and quantum-resistant authentication mechanisms were shown to significantly mitigate vulnerabilities, making BB84 applicable across diverse domains, including fiber, free-space, satellite, and underwater communication. At the same time, open challenges remain in distinguishing noise from malicious disturbances, achieving low-overhead error correction, extending finite-key security, and addressing side-channel vulnerabilities.

Looking ahead, the integration of BB84 with emerging technologies such as artificial intelligence, device-independent QKD, and global quantum networks is expected to drive its transition from experimental demonstrations to practical large-scale deployment. By addressing the identified challenges and adopting standardized frameworks, BB84 can serve as a foundation for secure communication infrastructures in the quantum era.

\section{Acknowledgments}

\textbf{\textit{Generative AI statement}}. The author(s) declare that the Generative AI tool ChatGPT was used to enhance the language and clarity of this work and take full responsibility for the accuracy and integrity of the content.

\bibliographystyle{IEEEtran}
\bibliography{references}

@article{Gisin_2002,
  author = {Gisin, Nicolas and Ribordy, Gr\'egoire and Tittel, Wolfgang and Zbinden, Hugo},
  title = {Quantum cryptography},
  journal = {Rev. Mod. Phys.},
  volume = {74},
  issue = {1},
  pages = {145--195},
  numpages = {0},
  year = {2002},
  month = {Mar},
  publisher = {American Physical Society},
  doi = {10.1103/RevModPhys.74.145},
  url = {https://link.aps.org/doi/10.1103/RevModPhys.74.145}
}

@article{Scarani_2009,
  author = {Scarani, Valerio and Bechmann-Pasquinucci, Helle and Cerf, Nicolas J. and Du\ifmmode \check{s}\else \v{s}\fi{}ek, Miloslav and L\"utkenhaus, Norbert and Peev, Momtchil},
  title = {The security of practical quantum key distribution},
  journal = {Rev. Mod. Phys.},
  volume = {81},
  issue = {3},
  pages = {1301--1350},
  numpages = {0},
  year = {2009},
  month = {Sep},
  publisher = {American Physical Society},
  doi = {10.1103/RevModPhys.81.1301},
  url = {https://link.aps.org/doi/10.1103/RevModPhys.81.1301}
}

@article{Bennett_2014,
author = {Charles H. Bennett and Gilles Brassard},
title = {Quantum cryptography: Public key distribution and coin tossing},
journal = {Theoretical Computer Science},
volume = {560},
pages = {7-11},
year = {2014},
note = {Theoretical Aspects of Quantum Cryptography – celebrating 30 years of BB84},
issn = {0304-3975},
doi = {https://doi.org/10.1016/j.tcs.2014.05.025},
url = {https://www.sciencedirect.com/science/article/pii/S0304397514004241}
}

@ARTICLE{Townsend_1998,
  author={Townsend, P.D.},
  journal={IEEE Photonics Technology Letters}, 
  title={Experimental investigation of the performance limits for first telecommunications-window quantum cryptography systems}, 
  year={1998},
  volume={10},
  number={7},
  pages={1048-1050},
  doi={10.1109/68.681313}
}

@article{Ursin_2007,
   author = {R. Ursin and F. Tiefenbacher and T. Schmitt-Manderbach and H. Weier and T. Scheidl and others},
   title={Entanglement-based quantum communication over 144 km},
   journal={Nature Physics},   
   volume={3},
   ISSN={1745-2481},
   url={http://dx.doi.org/10.1038/nphys629},
   DOI={10.1038/nphys629},
   number={7},
   publisher={Springer Science and Business Media LLC},
   year={2007},
   month=jun, pages={481–486} 
}

@article{Liao_2017,
  author    = {Sheng-Kai Liao and Wen-Qi Cai and Wei-Yue Liu and
            Liang Zhang and Yang Li and others},
  title     = {Satellite-to-ground quantum key distribution},
  journal   = {Nature},
  volume    = {549},
  number    = {7670},
  pages     = {43--47},
  year      = {2017},
  month     = sep,
  doi       = {10.1038/nature23655},
  url       = {https://doi.org/10.1038/nature23655},
  issn      = {1476-4687}
}

@article{Lo_1999,
author = {Hoi-Kwong Lo  and H. F. Chau },
title = {Unconditional Security of Quantum Key Distribution over Arbitrarily Long Distances},
journal = {Science},
volume = {283},
number = {5410},
pages = {2050-2056},
year = {1999},
doi = {10.1126/science.283.5410.2050},
URL = {https://www.science.org/doi/abs/10.1126/science.283.5410.2050},
eprint = {https://www.science.org/doi/pdf/10.1126/science.283.5410.2050},
}

@article{Mayers_2001,
author = {Mayers, Dominic},
title = {Unconditional security in quantum cryptography},
year = {2001},
issue_date = {May 2001},
publisher = {Association for Computing Machinery},
address = {New York, NY, USA},
volume = {48},
number = {3},
issn = {0004-5411},
url = {https://doi.org/10.1145/382780.382781},
doi = {10.1145/382780.382781},
journal = {J. ACM},
month = may,
pages = {351–406},
numpages = {56}
}

@Article{Fiorini_2024,
AUTHOR = {Fiorini, Francesco and Pagano, Michele and Garroppo, Rosario Giuseppe and Osele, Antonio},
TITLE = {Estimating Interception Density in the BB84 Protocol: A Study with a Noisy Quantum Simulator},
JOURNAL = {Future Internet},
VOLUME = {16},
YEAR = {2024},
NUMBER = {8},
ARTICLE-NUMBER = {275},
URL = {https://www.mdpi.com/1999-5903/16/8/275},
ISSN = {1999-5903},
DOI = {10.3390/fi16080275}
}

@Article{Termos_2024,
AUTHOR = {Termos, Hassan},
TITLE = {Quantum Authentication Evolution: Novel Approaches for Securing Quantum Key Distribution},
JOURNAL = {Entropy},
VOLUME = {26},
YEAR = {2024},
NUMBER = {6},
ARTICLE-NUMBER = {447},
URL = {https://www.mdpi.com/1099-4300/26/6/447},
PubMedID = {38920455},
ISSN = {1099-4300},
DOI = {10.3390/e26060447}
}

@article{Guitouni_2024,
    author = "Guitouni, Zied and Maize, Sirine and Zrigui, Mounir and Machhout, Mohsen",
    title = "{Advanced error correction method for quantum key distribution in IoT systems}",
    doi = "10.1088/1402-4896/ad7423",
    journal = "Phys. Scripta",
    volume = "99",
    number = "10",
    pages = "105106",
    year = "2024"
}

@article{Stucki_2002,
author = {Stucki, D and Gisin, N and Guinnard, O and Ribordy, G and Zbinden, H},
title = {Quantum key distribution over 67 km with a plug and play system},
journal = {New Journal of Physics},
doi = {10.1088/1367-2630/4/1/341},
url = {https://doi.org/10.1088/1367-2630/4/1/341},
year = {2002},
month = {jul},
publisher = {},
volume = {4},
number = {1},
pages = {41}
}

@article{Shor_2000,
  author = {Shor, Peter W. and Preskill, John},
  title = {Simple Proof of Security of the BB84 Quantum Key Distribution Protocol},  
  journal = {Phys. Rev. Lett.},
  volume = {85},
  issue = {2},
  pages = {441--444},
  numpages = {0},
  year = {2000},
  month = {Jul},
  publisher = {American Physical Society},
  doi = {10.1103/PhysRevLett.85.441},
  url = {https://link.aps.org/doi/10.1103/PhysRevLett.85.441}
}

@article{Wang_2013,
  author = {Jian-Yu Wang and Bin Yang and Sheng-Kai Liao and Liang Zhang and Qi Shen and others},
  title     = {Direct and full-scale experimental verifications towards ground--satellite quantum key distribution},
  journal   = {Nature Photonics},
  volume    = {7},
  number    = {5},
  pages     = {387--393},
  year      = {2013},
  month     = may,
  doi       = {10.1038/nphoton.2013.89},
  url       = {https://doi.org/10.1038/nphoton.2013.89},
  issn      = {1749-4893}
}

@article{Bonato_2009,
author = {Bonato, C and Tomaello, A and Da Deppo, V and Naletto, G and Villoresi, P},
title = {Feasibility of satellite quantum key distribution},
journal = {New Journal of Physics},
doi = {10.1088/1367-2630/11/4/045017},
url = {https://doi.org/10.1088/1367-2630/11/4/045017},
year = {2009},
month = {apr},
publisher = {},
volume = {11},
number = {4},
pages = {045017}
}

@article{Brassard_2000,
  author = {Brassard, Gilles and L\"utkenhaus, Norbert and Mor, Tal and Sanders, Barry C.},
  title = {Limitations on Practical Quantum Cryptography},  
  journal = {Phys. Rev. Lett.},
  volume = {85},
  issue = {6},
  pages = {1330--1333},
  numpages = {0},
  year = {2000},
  month = {Aug},
  publisher = {American Physical Society},
  doi = {10.1103/PhysRevLett.85.1330},
  url = {https://link.aps.org/doi/10.1103/PhysRevLett.85.1330}
}

@article{Jain_2014,
author = {Jain, Nitin and Anisimova, Elena and Khan, Imran and Makarov, Vadim and Marquardt, Christoph and Leuchs, Gerd},
title = {Trojan-horse attacks threaten the security of practical quantum cryptography},
journal = {New Journal of Physics},
doi = {10.1088/1367-2630/16/12/123030},
url = {https://doi.org/10.1088/1367-2630/16/12/123030},
year = {2014},
month = {dec},
publisher = {IOP Publishing},
volume = {16},
number = {12},
pages = {123030}
}

@article{Gobby_2004,
    author = {Gobby, C. and Yuan, Z. L. and Shields, A. J.},
    title = {Quantum key distribution over 122 km of standard telecom fiber},
    journal = {Applied Physics Letters},
    volume = {84},
    number = {19},
    pages = {3762-3764},
    year = {2004},
    month = {05},
    issn = {0003-6951},
    doi = {10.1063/1.1738173},
    url = {https://doi.org/10.1063/1.1738173},
    eprint = {https://pubs.aip.org/aip/apl/article-pdf/84/19/3762/18588579/3762_1_online.pdf},
}

@article{Wilson_1927,
author = {Edwin B. Wilson},
title = {Probable Inference, the Law of Succession, and Statistical Inference},
journal = {Journal of the American Statistical Association},
volume = {22},
number = {158},
pages = {209--212},
year = {1927},
publisher = {Taylor \& Francis},
doi = {10.1080/01621459.1927.10502953},
URL = { https://www.tandfonline.com/doi/abs/10.1080/01621459.1927.10502953},
eprint = {https://www.tandfonline.com/doi/pdf/10.1080/01621459.1927.10502953}
}

@article{Clopper_1934,
    author = {CLOPPER, C. J. and PEARSON, E. S.},
    title = {THE USE OF CONFIDENCE OR FIDUCIAL LIMITS ILLUSTRATED IN THE CASE OF THE BINOMIAL},
    journal = {Biometrika},
    volume = {26},
    number = {4},
    pages = {404-413},
    year = {1934},
    month = {12},
    issn = {0006-3444},
    doi = {10.1093/biomet/26.4.404},
    url = {https://doi.org/10.1093/biomet/26.4.404},
    eprint = {https://academic.oup.com/biomet/article-pdf/26/4/404/823407/26-4-404.pdf},
}

@article{Renner_2008,
author = {RENNER, RENATO},
title = {SECURITY OF QUANTUM KEY DISTRIBUTION},
journal = {International Journal of Quantum Information},
volume = {06},
number = {01},
pages = {1-127},
year = {2008},
doi = {10.1142/S0219749908003256},
URL = { https://doi.org/10.1142/S0219749908003256},
eprint = { https://doi.org/10.1142/S0219749908003256}
}

@article{Korzh_2015,
  author    = {Boris Korzh and Charles Ci Wen Lim and Raphael Houlmann and Nicolas Gisin and Ming Jun Li and Daniel Nolan and Bruno Sanguinetti and Rob Thew and Hugo Zbinden},
  title     = {Provably secure and practical quantum key distribution over 307 km of optical fibre},
  journal   = {Nature Photonics},
  volume    = {9},
  number    = {3},
  pages     = {163--168},
  year      = {2015},
  month     = mar,
  doi       = {10.1038/nphoton.2014.327},
  url       = {https://doi.org/10.1038/nphoton.2014.327},
  issn      = {1749-4893}
}

@article{Yin_2017,
author = {Juan Yin and Yuan Cao and Yu-Huai Li and
          Sheng-Kai Liao and Liang Zhang and others},
title = {Satellite-based entanglement distribution over 1200 kilometers},
journal = {Science},
volume = {356},
number = {6343},
pages = {1140-1144},
year = {2017},
doi = {10.1126/science.aan3211},
URL = {https://www.science.org/doi/abs/10.1126/science.aan3211},
eprint = {https://www.science.org/doi/pdf/10.1126/science.aan3211},
}

@article{Tomaello_2011,
author = {Andrea Tomaello and Cristian Bonato and Vania {Da Deppo} and Giampiero Naletto and Paolo Villoresi},
title = {Link budget and background noise for satellite quantum key distribution},
journal = {Advances in Space Research},
volume = {47},
number = {5},
pages = {802-810},
year = {2011},
note = {Scientific applications of Galileo and other Global Navigation Satellite Systems - II},
issn = {0273-1177},
doi = {https://doi.org/10.1016/j.asr.2010.11.009},
url = {https://www.sciencedirect.com/science/article/pii/S0273117710007362}
}

@article{Lo_2005,
  author = {Lo, Hoi-Kwong and Ma, Xiongfeng and Chen, Kai},
  title = {Decoy State Quantum Key Distribution},
  journal = {Phys. Rev. Lett.},
  volume = {94},
  issue = {23},
  pages = {230504},
  numpages = {4},
  year = {2005},
  month = {Jun},
  publisher = {American Physical Society},
  doi = {10.1103/PhysRevLett.94.230504},
  url = {https://link.aps.org/doi/10.1103/PhysRevLett.94.230504}
}

@article{Lucamarini_2018,
  author    = {M. Lucamarini and Z. L. Yuan and J. F. Dynes and A. J. Shields},
  title     = {Overcoming the rate--distance limit of quantum key distribution without quantum repeaters},
  journal   = {Nature},
  volume    = {557},
  number    = {7705},
  pages     = {400--403},
  year      = {2018},
  month     = may,
  doi       = {10.1038/s41586-018-0066-6},
  url       = {https://doi.org/10.1038/s41586-018-0066-6},
  issn      = {1476-4687}
}

@misc{Elkouss_2011,
      author={David Elkouss and Jesus Martinez-Mateo and Vicente Martin},
      title={Information Reconciliation for Quantum Key Distribution}, 
      year={2011},
      eprint={1007.1616},
      archivePrefix={arXiv},
      primaryClass={quant-ph},
      url={https://arxiv.org/abs/1007.1616}, 
}

@article{Tomamichel_2012,
  author    = {Marco Tomamichel and Charles Ci Wen Lim and Nicolas Gisin and Renato Renner},
  title     = {Tight finite-key analysis for quantum cryptography},
  journal   = {Nature Communications},
  volume    = {3},
  number    = {1},
  pages     = {634},
  year      = {2012},
  month     = jan,
  doi       = {10.1038/ncomms1631},
  url       = {https://doi.org/10.1038/ncomms1631},
  issn      = {2041-1723}
}

@article{Lydersen_2010,
  author    = {Lars Lydersen and Carlos Wiechers and Christoffer Wittmann and Dominique Elser and Johannes Skaar and Vadim Makarov},
  title     = {Hacking commercial quantum cryptography systems by tailored bright illumination},
  journal   = {Nature Photonics},
  volume    = {4},
  number    = {10},
  pages     = {686--689},
  year      = {2010},
  month     = oct,
  doi       = {10.1038/nphoton.2010.214},
  url       = {https://doi.org/10.1038/nphoton.2010.214},
  issn      = {1749-4893}
}

@article{Pirandola_2020,
author = {S. Pirandola and U. L. Andersen and L. Banchi and M. Berta and D. Bunandar and R. Colbeck and D. Englund and T. Gehring and C. Lupo and C. Ottaviani and J. L. Pereira and M. Razavi and J. Shamsul Shaari and M. Tomamichel and V. C. Usenko and G. Vallone and P. Villoresi and P. Wallden},
journal = {Adv. Opt. Photon.},
number = {4},
pages = {1012--1236},
publisher = {Optica Publishing Group},
title = {Advances in quantum cryptography},
volume = {12},
month = {Dec},
year = {2020},
url = {https://opg.optica.org/aop/abstract.cfm?URI=aop-12-4-1012},
doi = {10.1364/AOP.361502}
}

\end{document}